# Coreless 5G Mobile Network

Farooq Khan
Samsung Research America
Richardson, Texas, USA

*Abstract* — Today's mobile networks contain an increasing variety of proprietary hardware stifling innovation and leading to longer time-to-market for introduction of new services. We propose to replace the mobile core network nodes and interfaces with an Open Source SW implementation running on general purpose commodity hardware. The proposed open source approach referred to as coreless mobile network is expected to reduce cost, increase flexibility, improve innovation speed and accelerate time-to-market for introduction of new features and functionalities. A common Open Source SW framework will also enable automatic discovery and selection, seamless data mobility as well as unified charging and billing across cellular, Wi-Fi, UAV and satellite access networks.

## I. INTRODUCTION

Mobile connectivity has transformed daily life across the globe becoming one of the most dramatic game-changing technologies the world has ever seen. As more people connect to the Internet, increasingly chat to friends and family, watch videos on the move, and listen to streamed music on their mobile devices, mobile data traffic continues to grow following an omnify principle [1]. Here, *omnify* stands for *Order of Magnitude Increase every Five Years*. Which means, demand for data increases 10 times every 5 years and will continue to increase at this pace with expected 1,000 times increase in the next 15 years. This increase in demand is similar to the memory and computing power growth following Moore's law which offered a million fold more memory capacity and the processing power in the last 30 years. For wireless communications, it is more appropriate to measure advances in 5 years and 10 years timeframe as a new generation wireless technology is developed every 10 year and a major upgrade on each generation follows 5 years afterwards. Global mobile traffic already surpassed 1 Exabyte/ month mark in 2013 and is projected to grow 10 fold exceeding 10 Exabytes/ month within 5 years in 2018 [2]. With this trend, in 2028 global mobile traffic will cross the 1 Zettabyte/ month which is equivalent to 200 Gigabytes/month for 5 Billion users. Wi-Fi offload accounts for almost as much traffic as on mobile networks and also follows a similar growth trend with another Zettabyte/ month mobile traffic flowing over Wi-Fi networks by 2028.

In order to address the continuously growing wireless capacity challenge, the author and his colleagues pioneered use of higher frequencies referred to as millimeter waves with a potential availability of over 100GHz spectrum for 5G mobile communications [3]-[6]. At millimeter wave frequencies, radio spectrum use is lighter and very wide bandwidths along with a large number of smaller size antennas can be used to provide orders of magnitude increase in capacity needed in the next 15 to 20 years.

We will also need to complement cellular and Wi-Fi networks with satellites and other aerial systems such as those employing unmanned aerial vehicles (UAVs) [7]. More recently, the use of advanced consumer electronics technologies and adapting them to use in satellites and UAVs have enabled small platforms at increasingly lower cost, size and power consumption. Similarly spacecraft launch costs are experiencing downward pressure thanks to the entry of many private companies into the space race.

A constellation of satellites can cover virtually all of the inhabited Earth's surface and more. Even a few satellites can cover a much vaster number of potential subscribers than any terrestrial network. With transition of cellular to 5G millimeter wave spectrum, we can develop a single wireless technology for satellites, cellular and Wi-Fi access as well as for back-haul communications to take advantage of the scale and further reduce costs to make affordable Internet services available to everyone in the world.

Another barrier to reduce cost for mobile communications is the mobile packet core network which is plagued by continuously complexity requiring an increasing variety of proprietary hardware. The complex nature of the core network stems from many different network nodes, protocols and interfaces which add unnecessary overhead and delays undermining performance of mobile applications. Therefore, mobile networking industry is under immense pressure to innovate, and create agile, scalable and cost effective networks. In this paper, we propose a new coreless mobile network approach which leverages Open Source SW running on commodity hardware to provide a scalable, reliable and affordable solution that seamlessly supports cellular, Wi-Fi, Satellite and UAV based wireless access.

## II. MOBILE NETWORK ARCHITECTURE

In order to greatly simplify the wireless access, we propose getting rid of the core network nodes and interfaces and replacing them with Open Source SW implementation as depicted in Figure 1. All the necessary functionalities of a mobile network such as access control, mobility management, charging and radio resource control etc. are implemented as application programs (virtual network functions) build on top of the Open Source SW. In addition to benefits such as reduced cost and increased flexibility, open source approach improves innovation speed accelerating time-to-market for introduction of new services. A common Open Source SW framework will also allow automatic discovery and selection, seamless data mobility as well as unified charging and billing across Wi-Fi, cellular, UAV and satellite access networks.

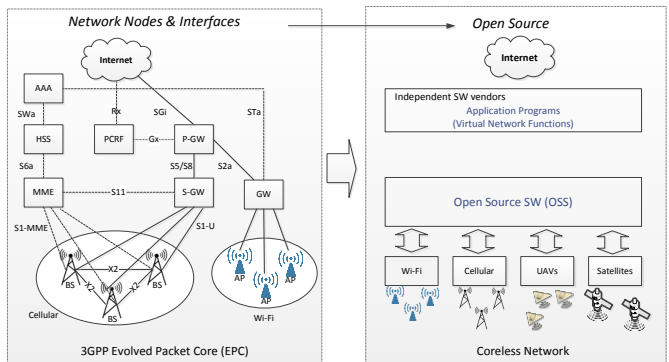

**Figure 1 Transition to Coreless Mobile network**

Let us first review a few functionalities and services offered by the 3GPP LTE Evolved Packet Core (EPC). In later sections, we will discuss how we can deploy virtual versions of network functions as well as transition to a coreless network.

### A. LTE Evolved Packet Core

The 4G LTE architecture defines the Evolved Packet System (EPS) as a combination of the LTE access system (radio part) and an Internet Protocol (IP) -based core network, the Evolved Packet Core (EPC). The 2G and 3G network architectures were designed to process voice and data via two separate sub-domains: circuit-switched for voice and packet-switched for data. The EPC unifies voice and data on an IP service architecture and voice is treated as just another IP application. The EPC also separates the user data (also known as the user plane) and the signaling (also known as the control plane) to make the scaling independent. The key components of the EPC are:

*Gateways (GW)*: The gateways (Serving GW and PDN GW) are user-plane nodes providing data paths between eNodeBs and the external IP networks. One of the essential functionality of the SGW, beside routing and forwarding packets, is as a local mobility anchor point for inter-eNodeB handovers as well as managing mobility between LTE and 2G/GSM and 3G/UMTS networks. The SGW is logically connected to the other gateway, the PDN GW (PGW) which acts as the interface between the LTE network and other IP networks. The PGW is responsible for IP address allocation for the UE, QoS enforcement, deep packet inspection (DPI) and flow-based charging according to rules from the PCRF (Policy Control and Charging Rules Function). The PGW also acts as a mobility "anchor" between 3GPP and non-3GPP technologies such as Wi-Fi as depicted in Figure 1. The 3GPP standard specifies these gateways independently but in practice they may be combined in a single "box" by network vendors.

*MME (Mobility Management Entity)*: The MME deals with the control plane and manages session states as well as authenticates and tracks a user across the network.

*Policy Control and Charging Rules Function (PCRF)*: The PCRF is responsible for policy control decision-making, as well as for controlling the flow-based charging functionalities in the Policy Control Enforcement Function (PCEF), which resides in the P-GW. The PCRF provides the QoS authorization that decides how a certain data flow will be treated in the PCEF and ensures that this is in accordance with the user's subscription profile.

*Home Subscriber Server (HSS)*: The HSS is basically a database that contains user information. It also provides support functions in mobility management, call and session setup, user authentication and access authorization.

In addition to the network nodes and functions defined in the EPC standard, today's networks need to implement many other functions such as firewalls, NATs, web security, virtual private network gateways, load balancers, DNS server, deep-packet inspectors, intrusion-detection systems, and WAN accelerators etc. The traditional 'function-per-box' approach would require proprietary hardware and software to implement and deploy these functions.

### B. Traffic Offloading

Mobile offloading has become extremely important with the surge of mobile traffic and devices because it enables mobile operators to optimize RAN resources, and improve the quality of experience (QoE) for data-intensive mobile applications. Since the first introduction of EPC in 3GPP Release 8 standard, more features have been added in later releases of the standard, related to traffic offloading. For example, Release 10 specifications introduced local breakout and latency reduction functions such as Local IP Access (LIPA) and Selected Internet IP Traffic Offload (SIPTO). Both techniques reduce traffic loads on EPC by introducing L-GWs (local gateways) which offload traffic directly to the Internet bypassing the core network elements.

In order to reduce traffic load on the radio access as well, a new technique referred to as IP Flow Mobility (IFOM) is defined as part of the 3GPP Release 12 standard. The IFOM enhances the WiFi offload capability by allowing simultaneous connection to both WiFi and LTE network. In earlier interworking scenarios UE was forced to use/select one radio network and make a selection to move to an alternative radio for all its traffic. In the IFOM architecture, there is no 'break out' point within the access network, where the IP data is separated. Instead, the terminal itself splits data traffic appropriately.

Another initiative related to traffic offloading is Hot Spot 2.0 (HS 2.0), also called Wi-Fi Certified Passpoint™, which is a new standard for public-access Wi-Fi that enables seamless roaming among WiFi networks and between WiFi and cellular networks. The goal is to enable users to automatically and securely connect to Wi-Fi networks while on the move providing an experience similar to the cellular systems. HS 2.0 was developed by the Wi-Fi Alliance and the Wireless Broadband Association to enable seamless hand-off of traffic without requiring additional user sign-on and authentication. HS 2.0 automates network discovery, registration and provisioning by leveraging three technologies: *WPA2*, *EAP* and *802.11u*.

*Wi-Fi Protected Access II (WPA2)*: *The WPA2* is security protocol and security certification program developed by the Wi-Fi Alliance. WPA2 implements the mandatory elements of IEEE 802.11i standard and has been in use since 2004, when it became available to ensure mutual authentication and encryption between the mobile device and network. WPA2 air-interface encryption is designed to make the Wi-Fi radio link as secure as in cellular.

*Extensible Authentication Protocol (EAP)*: The EAP is an authentication framework specified by IETF in RFC 3748. EAP defines how security credentials are moved between a mobile device and the security server. Hotspot 2.0 supports both EAP-SIM and EAP-TTLS based authentications with the goal to make the authentication as secure and seamless as in cellular systems. EAP-SIM allows the SIM card on a smartphone to be used for authentication on Wi-Fi networks. In EAP-TTLS (Tunneled Transport Layer Security), a username and password is assigned to mobile devices, such as tablets and laptops that do not have SIM cards. With EAP-TTLS, user credentials are transported in a securely encrypted tunnel established based upon the server certificates without the need for issuing certificates to users.

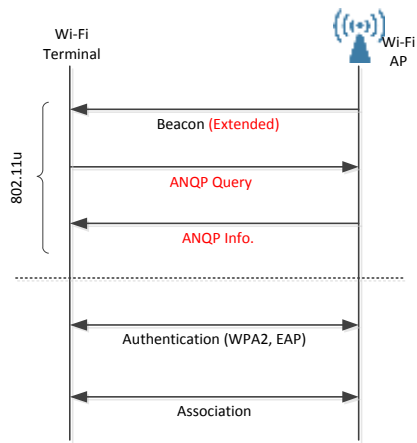

**Figure 2 Hotspot 2.0/ 802.11u**

*IEEE 802.11u*: This standard published in 2011 specifies techniques that allow a mobile device to collect information from a Wi-Fi network before association and authentication. 802.11u also enables Wi-Fi hotspots to advertise their capabilities and then allows devices to connect to them automatically rather than requiring the end user to manually select an SSID. The goal is to enable Wi-Fi connections as easily as cellular connections during roaming. A query and response protocol called Access Network Query Protocol (ANQP) forms the basis of 802.11u. The ANQP communicates information such as the AP operator's domain name, the IP addresses available at the AP, EAP method supported for authentication, and information about potential roaming partners accessible through the AP. An AP that is compliant with Wi-Fi Hotspot 2.0 begins the interaction by sending extended beacons as depicted in Figure 2. Using the ANQP, the Wi-Fi terminal sends a request that inquires which roaming consortium it belongs to, which network authentication to use, a Network Address Identifier (NAI) realm list etc. The terminal receives the response with these information elements and can then decide whether or not to connect.

We noted that with initiatives such as HS 2.0, the goal is to provide a Wi-Fi experience similar to the cellular systems in particular for support of roaming and mobility.

Another protocol, recently standardized by IETF, called multipath TCP (MP-TCP) [8] let mobile devices send and receive data across different network paths and interfaces such as cellular and Wi-Fi at the same time. In addition to benefits such as traffic offloading, higher data rates by simultaneously using more than one interface, MP-TCP provides improved network utilization, higher throughput, and greater resiliency by letting the network automatically and smoothly react to changing network conditions.

### III. NFV/ SDN

Today's mobile networks use 'function-per-box' approach containing an increasing variety of proprietary hardware. Each network appliance is slightly different in that it is optimized to support its particular function. Launching new network services frequently requires adding a new appliance. Many service providers believe that this 'function-per-box' model constrains the innovation and deployment of new network services and reduces ROI.

The concept of Network Function Virtualization (NFV) originated out of this frustration as service providers were finding it more and more difficult to deploy new network services to support their revenue and growth objectives [9]. NFV leverages standard IT virtualization technology to consolidate many network equipment types on to industry standard high volume servers, switches and storage or even on cloud computing infrastructure. Therefore, network functions that previously required a dedicated box can be replaced with virtual appliances running on say commodity x86 hardware as shown in Figure 3. Virtual appliances, like virtual machines, incorporate an application, OS and virtual hardware. Another alternative is to use LinuX Containers (LXC) which is an operating system-level virtualization method for running multiple isolated Linux systems (containers) on a single control host (LXC host). It does not provide a virtual machine, but rather provides a virtual environment that has its own CPU, memory, block I/O, network, etc. space.

The ability to deploy virtual versions of network functions on standard hardware anywhere in the network eliminates the need to install new equipment and greatly simplifies network management. The NFV not only reduce CAPEX and OPEX through reduced equipment costs and power consumption but also complexity and makes managing a network and deploying new capabilities easier and faster. Furthermore, NFV allows scaling up or down services to address changing demands without changing the underlying hardware. For example, instead of deploying a new hardware appliance to enable network encryption, NFV apps can be deployed using virtualized appliances on standardized servers or switches already in the network.

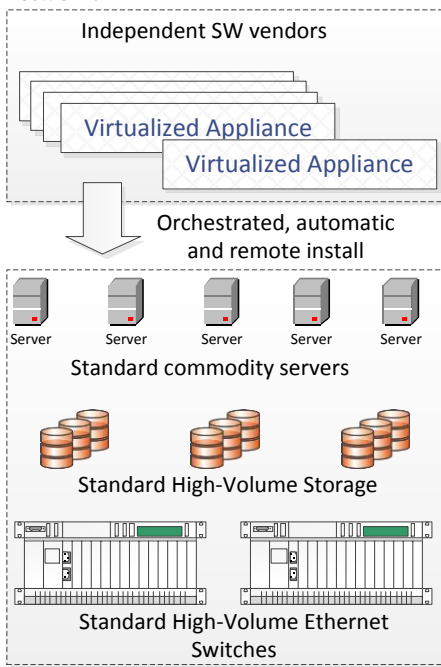

**Figure 3 Network Function Virtualization (NFV)**

Whereas NFV was created by a consortium of service providers, a new networking paradigm referred to as SDN (Software-defined Networking) got its start on campus networks [10]. The SDN separates the network's control (brains) and forwarding (muscle) planes and provides a centralized view of the distributed network for more efficient orchestration and automation of network services. In a classical router or switch architecture, the packet forwarding (data path) and the high level routing decisions (control path) occur on the same device [11]. The routing protocol engine programs forwarding decisions on the local device, i.e. router, using OSPF or BGP. In SDN, the data path portion still resides on the switch, while high-level routing decisions are moved to a centralized controller, typically a standard server. One of the first areas of application and success of SDN was in cloud data centers.

OpenFlow protocol, defined by The Open Networking Forum (ONF), is considered one of the first SDN standards. OpenFlow allows an SDN Controller to tell network switches where to send packets. In a conventional network, each switch has proprietary software that tells it what to do. With OpenFlow, the packet-moving decisions are centralized, so that the network can be programmed independently of the individual switches and routers. In addition to ONF, the OpenDaylight Project also aims to advance open standards and SDN adoption through the creation of a common industry supported platform.

We summarize the key attributes of NFV and SDN in Table 1. It should be noted that NFV is not dependent on SDN or SDN concepts. It is entirely possible to implement a virtualized network function (VNF) as a standalone entity using existing networking and orchestration paradigms. However, there are inherent benefits in leveraging SDN concepts to implement and manage an NFV infrastructure, particularly when looking at the management and orchestration of VNFs. A central orchestration and management system can take service provider requests associated with a VNF and translate them into the appropriate processing, storage and network configuration needed to bring the VNF into operation. Once in operation, the VNF potentially needs to be monitored for capacity and utilization and adapted if necessary. All these functions can be accomplished using SDN concepts and NFV could be considered one of the primary SDN use cases in service provider environments. Further, we note that SDN is focused on optimizing the underlying network infrastructure such as Ethernet switches and routers. On the other hand, the NFV decouples network elements from underlying hardware and aims to optimize deployment of network functions. Therefore, in principle, NFV can operate along with SDN taking advantage of optimized network infrastructure enabled by SDN.

**Table 1. NFV vs. SDN**

| NFV | SDN |
|---|---|
| Created by service providers | Born in the campus, matured in the data center |
| Decouple network elements from underlying hardware | Decouple control plane from data plane |
| Commoditize the telco specific hardware | Commoditize routers and switches |
| Optimize deployment of network functions such as load balancer, firewall, deep packet inspection etc. | Optimize network infrastructure such as Ethernet switches and routers |

## IV. CORELESS MOBILE NETWORK

Our vision of a coreless mobile network for 5G turns network elements and functions into software applications using principles of SDN/ NFV as depicted in Figure 4. We apply these principles to not just core network elements and functions but also to radio access network elements and functions such as handoffs and Radio Resource Management (RRM) etc. Here, we use the term WAP to represent a cellular base station, a Wi-Fi access point, a UAV, a satellite providing access or backhaul. With this new coreless network approach, all the network nodes, elements and interfaces of a traditional wireless network are replaced with programs built on open source code. A Coreless Mobile Network (CoMN) controller relays information to switches/routers and WAPs 'below' (via southbound APIs) and the applications 'above' (via northbound APIs).

The SDN network created is further virtualized using network virtualization (NV) techniques which creates a logical, virtual network, by decoupling network functions from the hardware that deliver them. All network functionality is separated from the underlying hardware and simulated as a "virtual instance" that can be loaded onto general, off-the-shelf platforms; a single hardware platform can be used to support multiple virtual network instances. We further leverage standard IT virtualization technology to virtualize the computing and storage hardware creating logical virtual compute and storage resources. We can then implement all the network and radio resource control functions on top of this virtual infrastructure.

Traditional mobile communication ecosystems were built by only 4 parties - equipment, device, chipset manufacturer and service provider. With proposed coreless network approach, we expect that IT vendors, open source communities and software developers will also be involved in the process. Open source communities can create the software code or hardware necessary to implement coreless approach based on common industry requirements. In this case, an open source community shares the burden of integration, testing and validation of the solutions, which results in increased industry-wide interoperability, reliability and efficiency that can shorten time to market compared to lengthy and complex process of developing and implementing industry standards.

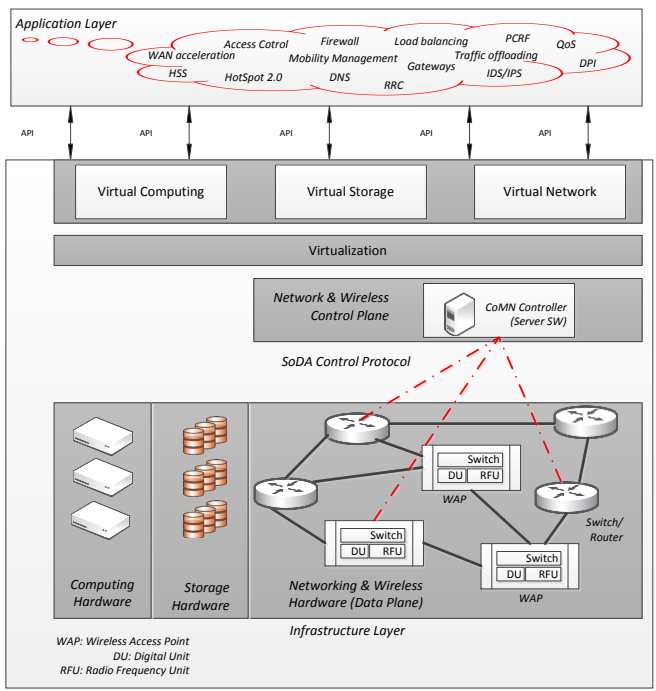

**Figure 4 Coreless Mobile network**

In the proposed approach, Industry leaders, academics and open source communities can come together to develop requirements and possibly a simple specification for a coreless mobile network. An open source software reference implementation for the coreless network can then be developed, for example, under Linux Foundation.

Conventional base stations using unified RF units (RFU) and baseband processing units (BBU) located at the base station site are considered workhorse of today's wireless networks. Most recently, a new architecture referred to as Centralized RAN or C-RAN is gaining traction with service providers. In C-RAN, the bulk of baseband processing is centralized and aggregated for a large number of distributed radio nodes. In comparison to standalone clusters of base stations, C-RAN provides significant performance and economic benefits such as baseband pooling, enhanced coordination between cells, virtualization, network extensibility, smaller deployment footprint and reduced power consumption.

In the C-RAN architecture, the radio function unit, also referred to as the radio frequency unit (RFU), is separated from the digital function unit or the baseband unit (BBU) by fiber. The digital baseband signals are carried over the fiber, usually using OBSAI or CPRI standard. The deployments of C-RAN so far have been restricted to the advanced wireless markets of Asia such as Japan, Korea and China, where fiber to provide fronthaul (link between a BBU and an RFU) is abundant and high population densities make the architecture fiscally beneficial. However, in many parts of the World where fiber deployments are not ubiquitous, the cost of deploying new fiber breaks the business case for service providers without existing fiber assets. The second challenge centers on virtualization of the physical layer which involves real-time processes and high computational load functions. General purpose processors are less efficient in running these signal processing functions and are not as optimized as dedicated SoC platforms when it comes to executing intensive physical layer tasks, such as channel decoding, FFT, and large-scale MIMO decoding. Dedicated SoC platforms can have as much as 10x the performance per Watt of general purpose processors for signal processing intensive physical layer functions.

We propose a new access architecture that we refer to Software Defined Access (SoDA) that can overcome the challenges of conventional C-RAN while still providing the benefit of centralized control and coordination as depicted in Figure 5. By applying principles of SDN to radio access network, we separate the radio control and data transmission planes with data transmission portion (BBU+RFU) residing on the Wireless Access Point (WAP), while high-level radio resource management (RRM) decisions moved to a separate centralized SoDA controller, typically a standard server. The SoDA controller with centralized Intelligence uses standard API to program the baseband unit and the RF unit in the WAPs. This greatly simplifies configuration as the entire radio access network, often comprised of thousands of WAPs from different vendors and of different types (cellular, Wi-Fi, UAV, satellite) can be programmed with a single API. Moreover, any RRM upgrades can be achieved independently from the WAP hardware.

The logically centralized control layer enables radio resource allocation decisions to be made with global visibility across many base stations, which is far more optimal than the distributed radio resource management, mobility management, and routing applications/protocols in in traditional vertically integrated approach. SoDA controller can change or update the scheduling algorithm and other physical layer parameters in the WAPs based on the global view of network and traffic conditions. For example, it can configure how many and which comm-cores should be active in a WAP supporting multi-comm-core architecture [1]. In addition, scalability is improved because as new users are added, the required compute capacity at each base station remains low because RRM processing is centralized in the SoDA controller.

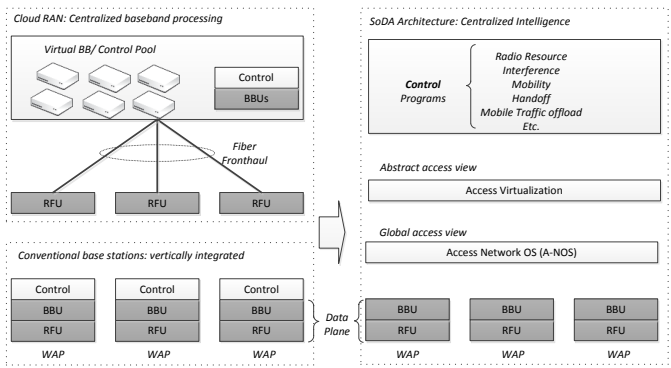

**Figure 5 Software Defined Access (SoDA) Architecture**

The fronthaul capacity needed in C-RAN is generally an order of magnitude higher compared to the actual information rate because it needs to carry digital baseband signals. The data compression techniques can be used to compress the digital baseband signals at the expense of additional complexity and signal delays. A single 20 MHz 2x2 MIMO LTE channel requires about 2.5 Gb/s fronthaul capacity. With use of very large bandwidths approaching GHz and massive MIMO in 5G, we expect fronthaul load to increase by hundred times or more for a C-RAN based deployment. This unsustainable level of fronthaul capacity may make C-RAN practically infeasible for 5G from cost perspective. In the proposed SoDA architecture, there is no need for the fronthaul as both BBU and RFU are collocated in the WAP. In this case, the backhaul network connecting BBU or digital unit (DU) with wireless network routers and IP networks just carries information packets (not digital baseband signals) and SoDA control protocol information.

The SoDA architecture is also suitable for 5G millimeter wave systems where small size of antennas and other RF components enables BBU and RFU integration leading to small form factor WAPs. These compact WAPs can be mounted outdoor or indoor - on towers, on utility poles, sides of buildings or anywhere a power connection exist, making installation less costly and easier. Therefore, SoDA architecture offers the benefits of C-RAN without the need for a fronthaul.

## V. CONCLUSION

We proposed an open source software approach for a coreless mobile network with a goal to reduce cost, eliminate vendor lock-in, improve innovation speed and accelerate time-to-market for deployment of new services. The proposed coreless network approach based on open source turns both core and radio access network functions into software programs eliminating the need to standardize an ever increasing number of network nodes and interfaces. With the proposed approach, new network features and functions can be introduced in a speedy manner without going through lengthy and complex process of developing "standard-per-function". Moreover, network functions can easily be programmed to suit and seamlessly support different types of wireless access methods including those based on cellular, Wi-Fi, UAV and satellites communication. We expect that IT vendors, open source communities and software developers in addition to traditional players such as equipment, device, chipset manufacturers and service providers will need to be involved to make the coreless network approach and the open source eco system successful.

ACKNOWLEDGMENT

The author would like to thank his colleagues at SAMSUNG for valuable discussions and feedback.